\documentclass[]{aastex631}

\usepackage{amsmath}
\usepackage[ruled]{algorithm2e}
\usepackage{listings}
\usepackage{float}
\usepackage{longtable}
\usepackage{multirow} 
\usepackage{colortbl}

\newcommand{\vela}{\texttt{Vela.jl}}
\newcommand{\pint}{\texttt{PINT}}
\newcommand{\pyvela}{\texttt{pyvela}}
\newcommand{\enterprise}{\texttt{ENTERPRISE}}
\newcommand{\tempotwo}{\texttt{tempo2}}
\newcommand{\temponest}{\texttt{TEMPONEST}}
\newcommand{\multinest}{\texttt{MULTINEST}}
\newcommand{\polychord}{\texttt{PolyChord}}
\newcommand{\gu}{\texttt{GeometricUnits.jl}}
\newcommand{\emcee}{\texttt{emcee}}
\newcommand{\astropy}{\texttt{astropy}}
\newcommand{\toa}{\texttt{TOA}}
\newcommand{\comp}{\texttt{Component}}
\newcommand{\tm}{\texttt{TimingModel}}
\newcommand{\kernel}{\texttt{Kernel}}
\newcommand{\parf}{\texttt{par}}
\newcommand{\tim}{\texttt{tim}}
\newcommand{\spnta}{\texttt{SPNTA}}

\newcommand{\updated}[1]{#1}

\definecolor{codegreen}{rgb}{0,0.6,0}
\definecolor{codegray}{rgb}{0.5,0.5,0.5}
\definecolor{codepurple}{rgb}{0.58,0,0.82}
\definecolor{backcolour}{rgb}{0.95,0.95,0.92}
\lstdefinestyle{pystyle}{
  backgroundcolor=\color{backcolour}, commentstyle=\color{codegreen},
  keywordstyle=\color{magenta},
  numberstyle=\tiny\color{codegray},
  stringstyle=\color{codepurple},
  basicstyle=\ttfamily\footnotesize,
  breakatwhitespace=false,         
  breaklines=true,                 
  captionpos=b,                    
  keepspaces=true,                 
  numbers=left,                    
  numbersep=5pt,                  
  showspaces=false,                
  showstringspaces=false,
  showtabs=false,                  
  tabsize=2
}
\newfloat{lstfloat}{}{lop}
\floatname{lstfloat}{Code}

\begin{document}

\title{Bayesian pulsar timing and noise analysis with \vela{}: an overview}

\author[0000-0002-2820-0931]{Abhimanyu Susobhanan}
\affiliation{Max-Planck-Institut f{\"u}r Gravitationsphysik (Albert-Einstein-Institut), Leibniz Universit{\"a}t Hannover, Callinstra{\ss}e 38, 30167 Hannover, Deutschland}

\correspondingauthor{Abhimanyu Susobhanan}
\email{abhimanyu.susobhanan@aei.mpg.de}



\begin{abstract}

We present \vela{}, an efficient, modular, easy-to-use Bayesian pulsar timing and noise analysis package written in Julia.
\vela{} provides an independent, efficient, and parallelized implementation of the full non-linear pulsar timing and noise model along with a Python binding named \pyvela{}.
One-time operations such as data file input, clock corrections, and solar system ephemeris computations are performed by \pyvela{} with the help of the \pint{} pulsar timing package.
Its reliability is ensured via careful design utilizing Julia's type system, strict version control, and an exhaustive test suite.
This paper describes the design and usage of \vela{} focusing on the narrowband paradigm.

\end{abstract}

\keywords{Pulsars (1306) — Astronomy software (1855) — Astronomy data analysis (1858)}


\section{Introduction} \label{sec:intro}

Pulsars are rotating neutron stars whose electromagnetic radiation is received as periodic pulses.
Their high rotational stability makes them excellent celestial clocks, and pulsar timing, the technique of tracking a pulsar's rotation by measuring the times of arrival (TOAs) of its pulses, is one of the most precise techniques in astronomy \citep{LorimerKramer2012}. 
Over the years, pulsar timing has been applied to study a wide array of astrophysical phenomena, such as neutron star equation of state \updated{and internal dynamics} \citep[e.g.,][]{CromartieFonseca+2020,AntonelliMontoliPizzochero2022}, tests of theories of gravity \citep[e.g.,][]{VoisinCognard+2020}, solar wind \citep[e.g.,][]{TiburziShaifullah+2021}, galactic dynamics \citep[e.g.,][]{PereraBarr+2019}, etc.
Recently, pulsar timing array \citep[PTA:][]{FosterBacker1990} experiments found evidence \citep[e.g.,][]{AgazieAntoniadis+2024} for a stochastic gravitational wave background \citep[GWB:][]{HellingsDowns1983} in the nanohertz frequency range with the help of pulsar timing.

In addition to the pulsar rotation, the measured TOAs are also influenced by several deterministic astrophysical processes such as the orbital motion of the Earth, proper motion of the pulsar, solar wind, interstellar dispersion, the orbital motion of the pulsar, etc \citep{EdwardsHobbsManchester2006}, as well as stochastic processes such as radiometer noise, pulse jitter, rotational irregularities, interstellar medium variability, etc \citep{AgazieAnumarlapudi+2023_ng15detchar}. 
High-precision pulsar timing requires accurate modeling of all of these effects.
A pulsar timing model or pulsar ephemeris provides a mathematical description of the deterministic processes that affect the measured TOAs and is often accompanied by a noise model that describes the stochastic processes therein.

In practice, pulsar timing involves estimating the parameters of a pulsar timing model given a set of TOA measurements, usually in a frequentist setting using a software package such as \tempotwo{} \citep{HobbsEdwardsManchester2006, EdwardsHobbsManchester2006} or \pint{} \citep{LuoRansom+2021,SusobhananKaplan+2024}.
Noise characterization can be done in a few different ways.
In \pint{}, noise parameters can be estimated together with the timing parameters in a frequentist way \citep{SusobhananKaplan+2024}.
\tempotwo{} provides plugins like \texttt{fixData} and \texttt{spectralModel} that can estimate some noise parameters by fixing the timing model parameters \citep{Hobbs2014}.
Bayesian noise characterization can be performed using the  \enterprise{} \citep{JohnsonMeyers+2024} and \temponest{} \citep{LentatiHobson+2014} packages starting from a post-fit timing model, and is considered standard practice for high-precision timing experiments like PTAs. 
\enterprise{} and \temponest{} differ in how they treat the timing model fit; the former analytically marginalizes a linearized approximation of the timing model \citep{van_HaasterenLevin2013}, whereas the latter enables inference over the full non-linear timing and noise model.
The \pint{} package also supports Bayesian parameter estimation through \texttt{pint.fitter.MCMCFitter} and \texttt{pint.bayesian} interfaces although their use on large datasets is hampered by slow performance \citep{LuoRansom+2021,SusobhananKaplan+2024}.
Other packages that have been used for Bayesian pulsar timing and/or noise analysis include \texttt{libstempo} \citep{VigelandVallisneri2014}, \texttt{piccard} \citep{van_Haasteren2016},  \texttt{PAL2} \citep{EllisVan_Haasteren2017}, and \texttt{FortyTwo} \citep[e.g.][]{ChenCaballero+2021}.

In this paper, we present \vela{}\footnote{Named after the Vela pulsar \citep[J0835--4510:][]{LargeVaughanMills1968}, the brightest radio pulsar. Also, V{\=e}\b{l}a is a word meaning occasion, time, etc in Malayalam with cognates in several other Indian languages.}\footnote{The source code is available at \url{https://github.com/abhisrkckl/Vela.jl}. The documentation is available at \url{https://abhisrkckl.github.io/Vela.jl/}. \updated{This paper corresponds to version 0.0.7 \citep{Susobhanan2025}.}}, a new framework written in Julia \citep{BezansonEdelman+2017} for performing Bayesian inference over the full non-linear timing and noise model. 
\vela{} provides an efficient and modular implementation of the pulsar timing and noise model independent of other pulsar timing packages\footnote{Note that \vela{} relies on \pint{} to read input files, perform clock corrections, and compute solar system ephemerides. Aside from this, the timing \& noise model is implemented independently. See subsection \ref{sec:pyvela} for more details.}, and supports parallelized evaluation of the pulsar timing likelihood function using multi-threading.
We also provide a Python interface named \pyvela{} since Python is more popular among astronomers than Julia.
This package is developed with a focus on reliability, performance, and modularity (in that order), employing strict version control (using \texttt{git}) and rigorous testing.

The main objective of \vela{} is to provide a flexible, robust, modular, well-documented tool for performing Bayesian inference over the full non-linear pulsar timing and noise model.
While this functionality is already available in the \temponest{} package, \vela{} differs from \temponest{} in the following ways.
(1) \temponest{} can only be used with the nested sampling packages \multinest{} \citep{FerozHobson+2008} and \polychord{} \citep{HandleyHobson2015}, whereas \vela{} can be used with any Markov Chain Monte Carlo \citep[MCMC:][]{Diaconis2009} or nested sampler \citep{AshtonBernstein+2022} available in Julia or Python.
(2) \temponest{} uses \tempotwo{} internally to evaluate the timing model, whereas \vela{} contains an independent implementation of the timing model.
(3) \temponest{} only supports the narrowband timing paradigm, whereas \vela{} supports both the narrowband and wideband timing paradigms (the application of \vela{} on wideband data will be discussed in a separate paper). 
(4) \temponest{} is a stand-alone application whereas \vela{} is a library that can be used programmatically alongside various Julia and Python packages.

This paper is arranged as follows. 
In Section \ref{sec:timing-basics}, we provide a brief overview of the pulsar timing and noise analysis.
In Section \ref{sec:vela-overview}, we describe the design and implementation of \vela{} and \pyvela{}.
In Section \ref{sec:examples}, we provide three examples that showcase the functionalities of \vela{}.
Finally, in \ref{sec:summary}, we summarize our work and discuss the future directions.

\section{A brief overview of Bayesian pulsar timing and noise analysis}
\label{sec:timing-basics}

The fundamental measurable quantity in conventional pulsar timing is the TOA.
TOAs are measured by folding the time series pulsar data over a known pulsar period and cross-correlating the resulting integrated pulse profile against a noise-free template \citep{Taylor1992}, usually after splitting the observation into multiple frequency subbands (this is known as narrowband timing).
The more recent wideband timing technique involves measuring a single TOA and a dispersion measure (DM) from an observation by matching the frequency-resolved integrated pulse profile against a two-dimensional template known as a portrait \citep{PennucciDemorestRansom2014, Pennucci2019}.
Pulsar timing can also be applied directly to integrated pulse profiles \citep[e.g.][]{LentatiHobson+2015}, although such techniques are unsuitable for longer pulsar timing campaigns due to intractable data volumes. 
In the case of high-energy observations, where it is difficult to form integrated pulse profiles or TOAs due to low photon counts, photon arrival times can be used directly for pulsar timing \citep[e.g.][]{AjelloAtwood+2022}.
In this paper, we focus exclusively on the narrowband timing paradigm.

A narrowband TOA $t_{\text{arr}}$ measured at a terrestrial observatory is related to the pulse emission time $t_{\text{em}}$ as
\begin{align}
    t_{\text{arr}} &= t_{\text{em}} + \Delta_{\text{B}}(t_{\text{em}}) + \Delta_{\text{ltt}} + \Delta_{\text{DM}}(t_{\text{em}},\nu) +  \Delta_{\text{scatter}}(t_{\text{em}},\nu) + \Delta_{\text{GW}}(t_{\text{em}}) + \Delta_{\odot}(t_{\text{em}})  \nonumber\\
    &\qquad + \Delta_{\text{clock}}(t_{\text{em}}) + \Delta_{\text{jump}} + \mathcal{N}_{\text{R}} + ... \,,
    \label{eq:delays}
\end{align}
where 
\updated{$\nu$ is the observing frequency,}
$\Delta_{\text{B}}$ represents the delays caused by the binary motion of the pulsar including R{\o}mer delay, Shapiro delay, and Einstein delay \citep{DamourDeruelle1986}, 
$\Delta_{\text{ltt}}$ represents a constant light travel time between the pulsar system barycenter and the solar system barycenter at some fiducial epoch, $\Delta_{\text{DM}}$ is the dispersion delay caused by free electrons in the interstellar medium and the solar wind \citep{BackerHellings1986}, 
$\Delta_{\text{scatter}}$ represents the delay caused by interstellar scattering \citep{HembergerStinebring2008},
$\Delta_{\text{GW}}$ represents gravitational wave (GW) induced perturbations to the light travel time \citep{EstabrookWahlquist1975}, 
$\Delta_{\odot}$ represents the delays caused by the motion of the Earth in the solar system including the R{\o}mer delay, Shapiro delay, and Einstein delay \citep{EdwardsHobbsManchester2006},  
$\Delta_{\text{clock}}$ is a series of corrections that converts the TOA measured against an observatory clock to a timescale defined at the solar system barycenter \citep{HobbsEdwardsManchester2006}, and
$\Delta_{\text{jump}}$ represents instrumental delays.
\updated{$\mathcal{N}_{\text{R}}$ is a stochastic term arising from the radiometer noise of the telescope \citep{LorimerKramer2012}.} 

Note that each of the delays $\Delta_{X}$ described above can have both deterministic and stochastic components, and such a separation is arbitrary and model-dependent in many cases. 
The stochastic components corresponding to the delay terms above are orbital variations (due to tidal effects, ablation/accretion from the companion, presence of a third body, etc) \citep[e.g.,][]{ArzoumanianFruchterTaylor1994}, dispersion and scattering variations (due to dynamic and inhomogeneous interstellar medium and variable solar wind) \citep[e.g.,][]{KrishnakumarMitra+2015,KrishnakumarManoharan+2021}, the stochastic GW background \citep[e.g.,][]{AgazieAnumarlapudi+2023_gwb}, solar system ephemeris errors \citep[e.g.,][]{VallisneriTaylor+2020}, clock errors \citep[e.g.,][]{TiburziHobbs+2015}, etc.
Some of these processes, such as stochastic GW background, solar system ephemeris errors, and clock errors, are correlated across multiple pulsars.
Such processes usually cannot be distinguished from the rotational irregularities of the pulsar (described below) \updated{when only one pulsar is considered}, and are therefore not included in single-pulsar analyses.

The emission time $t_{\text{em}}$ estimated using equation \eqref{eq:delays} can be related to the pulsar rotational phase $\phi$ using the equation \citep{HobbsEdwardsManchester2006}
\begin{equation}
    \phi = \phi_0 + \sum_{n=0}^{N_F} \frac{F_n}{(n+1)!}  (t_{\text{em}}-t_0)^{n+1} + \phi_{\text{glitch}}(t_\text{em}) + \phi_{\text{SN}}(t_\text{em}) + \phi_{\text{prof}}(t_\text{em},\nu) + \mathcal{N}_{\text{jitter}} + ...\,,
    \label{eq:phasing}
\end{equation}
where $\phi_0$ represents an arbitrary initial phase, $F_n$ represent the pulsar frequency and its derivatives, $N_F$ is the number of frequency derivatives, $\phi_{\text{glitch}}$ represents phase corrections due to glitches, and $\phi_{\text{SN}}$ is a stochastic term representing slow stochastic variations in the pulsar rotation known as spin noise.
\updated{Here, we have defined $\phi$ such that one full rotation corresponds to $\phi=1$.} 
The initial phase $\phi_0$  can only be measured modulo an integer number of full rotations, and the constant light travel time $\Delta_{\text{ltt}}$ is fully covariant with $\phi_0$ and it is therefore excluded from equation \eqref{eq:delays} in practice.
\updated{Effects that alter the shape of the integrated pulse profile, such as frequency-dependent profile evolution \citep{HankinsRickett1986}, profile shape change events \citep[e.g.][]{SinghaSurnis+2021}, etc, thought to be of pulsar magnetospheric origin, can also influence the computed pulse phase although they are not directly related to pulsar rotation.
These are represented together as $\phi_{\text{prof}}$ in the above equation.
Similarly, random pulse shape variations \citep[pulse jitter;][]{ParthasarathyBailes+2021} are represented using the  stochastic term $\mathcal{N}_{\text{jitter}}$.}

Further, if a delay term $\Delta_{X}$ is sufficiently small compared to the pulse period $F_0^{-1}$, it can be moved into equation \eqref{eq:phasing} as a phase $\Delta_{X} F_0$ (e.g., $\Delta_{\text{jump}}$), and if a phase $|\phi_X| \ll 1$, it can be moved into equation \eqref{eq:delays} as a delay $\phi_X / F_0$ (e.g., $\phi_{\text{SN}}$).
\updated{We have arranged equations \eqref{eq:delays} and \eqref{eq:phasing} such that equation \eqref{eq:delays} contains effects that are external to the pulsar whereas equation \eqref{eq:phasing} contains pulsar-intrinsic effects.
However, their implementation follows historical conventions to be consistent with other pulsar timing packages.}

The timing residual $r$ is defined as 
\begin{equation}
    r = \frac{\phi - \mathfrak{N}[\phi]}{F}\,,
    \label{eq:resids}
\end{equation}
where $\mathfrak{N}[\phi]$ is the integer closest to $\phi$ and $F={d\phi}/{dt_{\text{arr}}}$ is the topocentric pulse frequency \citep{HobbsEdwardsManchester2006}.

~

The pulsar timing log-likelihood function $\ln L$ can be written as
\begin{equation}
    \ln L = -\frac{1}{2}\textbf{r}^T \textbf{C}^{-1} \textbf{r} - \frac{1}{2}\ln\det [2\pi\textbf{C}]\,,
    \label{eq:lnlike}
\end{equation}
where $\textbf{r}$ is an $N_{\text{toa}}$-dimensional column vector containing the timing residuals $r_i$ corresponding to the TOAs $t_i$, $\textbf{C}$ is the $N_{\text{toa}}\times N_{\text{toa}}$-dimensional TOA covariance matrix\footnote{\updated{The elements of \textbf{C} are defined as $C_{ij}=\left\langle t_{i} t_{j} \right\rangle - \left\langle t_{i}\right\rangle\left\langle t_{j} \right\rangle$ where $\left\langle \cdot \right\rangle$ represents ensemble average.}}, and $N_{\text{toa}}$ is the number of TOAs \citep{LentatiHobson+2014}.
Pulsar timing in a frequentist setting involves maximizing this likelihood over timing (and possibly noise) parameters \citep{HobbsEdwardsManchester2006,SusobhananKaplan+2024}.

The covariance matrix $\textbf{C}$ in general can be a dense symmetric matrix, \updated{and $\textbf{C}^{-1}$ and $\ln\det \textbf{C}$ can be computationally expensive to evaluate}.
To mitigate this computational cost, it is customary to use a reduced-rank approximation of $\textbf{C}$ where
\begin{equation}
    \textbf{C} = \textbf{N} + \textbf{U}^T\boldsymbol{\Phi}\textbf{U}\,,
\end{equation}
such that $\textbf{C}^{-1}$ and $\ln\det \textbf{C}$ can be evaluated relatively inexpensively using the Woodbury lemma and the matrix determinant lemma \citep{van_HaasterenVallisneri2014a}.
Here, $\textbf{N}$ is an $N_{\text{toa}}\times N_{\text{toa}}$-dimensional diagonal matrix containing scaled TOA measurement variances $\varsigma_i^2$ given by
\begin{equation}
    \varsigma_i^2 = E_f^2 (\sigma_i^2 + E_q^2)\,,
    \label{eq:errcorr}
\end{equation}
where \updated{$\sigma_i$ is a TOA measurement uncertainty}, $E_f$ is an EFAC (`error factor'), and $E_q$ is an EQUAD (`error added in quadrature') \citep{LentatiHobson+2014}.
Physically, $\textbf{N}$ represents the fully uncorrelated noise present in the TOAs such as radiometer noise and the uncorrelated part of the pulse jitter noise which depend on the observing system.
$\textbf{U}$ is an $N_{\text{toa}}\times p$ dimensional rectangular `basis' matrix and $\boldsymbol{\Phi}$ is a $p\times p$ dimensional diagonal `weight' matrix where $N_{\text{toa}}$ is the number of TOAs and $p$ is the rank of $\textbf{C}$ such that $p\ll N_{\text{toa}}$.

~

The correlated stochastic processes present in the TOAs are often represented as Gaussian processes which can be written in the form of a delay $\boldsymbol{\Delta}_X = \textbf{U}_X^T \textbf{a}_X$, where $\textbf{U}_X$ is a `basis' matrix, $\textbf{a}_X$ is a vector of amplitudes which are a priori Gaussian-distributed with means $\left\langle\textbf{a}_X\right\rangle=0$ and (co)variances $\left\langle\textbf{a}_X \textbf{a}_X^T\right\rangle=\boldsymbol{\Phi}_X$, and $\boldsymbol{\Phi}_X$ is a diagonal `weight' matrix.
Such a process can be included in the timing \& noise analysis in two ways.
One, they can be included as a delay term in equation \eqref{eq:delays} along with hyperpriors imposed on the weights $\boldsymbol{\Phi}_X$.
Alternatively, if the contribution of $\boldsymbol{\Delta}_X$ to $\textbf{r}$ is \updated{approximately} linear in $\textbf{a}_X$ (which is valid in most cases), the amplitudes $\textbf{a}_X$ can be analytically marginalized assuming the above-mentioned Gaussian priors, and this moves the contribution of this process into  $\textbf{U}^T\boldsymbol{\Phi}\textbf{U}$, leaving only the weights $\boldsymbol{\Phi}_X$ as free parameters \citep{LentatiHobson+2014}.

For example, the spin noise is usually modeled as a Fourier series such that the elements of $\textbf{U}_{\text{SN}}$ are given by \citep{LentatiHobson+2014}
\begin{equation}
    U_{\text{SN};jk}=\begin{cases}
\sin\left[\frac{\pi(k+1)(t_{j}-t_{0})}{T_{\text{span}}}\right] & \text{for odd }k\\
\cos\left[\frac{\pi k(t_{j}-t_{0})}{T_{\text{span}}}\right] & \text{for even }k
\end{cases}\,,
\end{equation}
where $t_{0}$ is some fiducial epoch and $T_{\text{span}}$ is the total span of the TOAs. 
In this case, the amplitude vector $\textbf{a}_{\text{SN}}$ contains the corresponding Fourier coefficients and the weights contained in $\boldsymbol{\Phi}_{\text{SN}}$ can be interpreted as power spectral densities.
See Appendix \ref{sec:rednoise} for details on how such processes are implemented in \vela{}.
Another example is the ECORR, which represents the component of the pulse jitter noise that is correlated across narrowband TOAs derived from the same observation but uncorrelated otherwise; see Appendix \ref{sec:ecorr} for further details.

~

Finally, the log-posterior distribution $\ln P$ of the timing \& noise model parameters $\boldsymbol{\alpha}$ given a dataset $\mathfrak{D}$ (containing TOAs, TOA uncertainties, observing frequencies, instrumental configuration information, etc) and a timing \& noise model $\mathfrak{M}$ can be written using the Bayes theorem as
\begin{equation}
    \ln P[\boldsymbol{\alpha} | \mathfrak{D}, \mathfrak{M}] = 
    \ln L[\mathfrak{D} | \boldsymbol{\alpha}, \mathfrak{M}] 
    + \ln \Pi[\boldsymbol{\alpha} | \boldsymbol{A}, \mathfrak{M}] 
    + \ln \Pi[\boldsymbol{A} | \mathfrak{M}]
    - \ln Z[\mathfrak{D} | \mathfrak{M}]\,,
    \label{eq:lnpost}
\end{equation}
where $\Pi$ represents the prior distributions, $\boldsymbol{\alpha}$ represents the free parameters appearing in $\ln L$ (i.e., equations (\ref{eq:delays}--\ref{eq:lnlike})),  $\boldsymbol{A}$ are the hyperparameters that determine the prior distributions of $\boldsymbol{\alpha}$ (like the weights $\boldsymbol{\Phi}_X$), and $Z$ is a normalizing constant known as the Bayesian evidence.

~

The implementation of the pulsar timing \& noise model summarized above (given by equations (\ref{eq:delays}--\ref{eq:lnpost})) in \vela{} is detailed in the next section.

\section{The design and implementation of \vela{}}
\label{sec:vela-overview}

\subsection{Numerical precision}
Pulsar timing is one of the few applications where the required numerical precision exceeds the precision provided by the 64-bit floating point type available in most programming languages (e.g., \texttt{double} in C and C++, \texttt{float} in Python, \texttt{Float64} in Julia).
An extended precision floating point type is required to accurately represent the measured TOA values, the pulse phase, and in some cases, the pulsar rotational frequency.
\vela{} uses the \texttt{Double64} type provided by the \texttt{DoubleFloats.jl} package \citep{Sarnoff2022}, which represents an extended-precision number as a sum of two \texttt{Float64}s \citep{Dekker1971}, for this purpose.\footnote{The \tempotwo{} package (and by extension, \temponest{}) uses the \texttt{long double} type available in C++ to represent TOAs and other quantities internally.
Similarly, \pint{} uses the \texttt{numpy.longdouble} type provided by the \texttt{numpy} package for this purpose, which in turn is implemented using the C \texttt{long double}.
Unfortunately, \texttt{long double} is not fully defined by the C and C++ standards, and its properties are hardware and compiler-dependent.
In machines where \texttt{long double} does not provide adequate precision, \tempotwo{} falls back to the \texttt{\_\_float128} type available as a compiler extension in the GNU C Compiler (\texttt{gcc}).
\texttt{Double64} avoids this issue since it is defined in terms of the \texttt{Float64} type.}\footnote{Extended precision is not required to represent the linearized timing model such as in \enterprise{} since it is only concerned with small deviations from best-fit values.}
Note that only the TOA values and the rotational phases are represented using \texttt{Double64}, and all other quantities are represented using \texttt{Float64}s because the software-implemented \texttt{Double64} arithmetic is significantly slower than \texttt{Float64} arithmetic which is usually hardware-supported.

The rotational Frequency $F_0$ is handled manually as a sum of two \texttt{Float64} numbers as $F_0=F_0^\text{big} + F_0^\text{small}$, where $F_0^\text{big} \ll F_0^\text{small}$ and $F_0^\text{small}$ is treated as a free parameter where applicable.
Handling $F_0$ this way significantly simplifies the implementation by allowing all model parameters to have the same underlying floating point type.

\begin{figure}[t]
    \centering
    \includegraphics[width=0.7\linewidth,trim={0 5.3cm 0 0},clip]{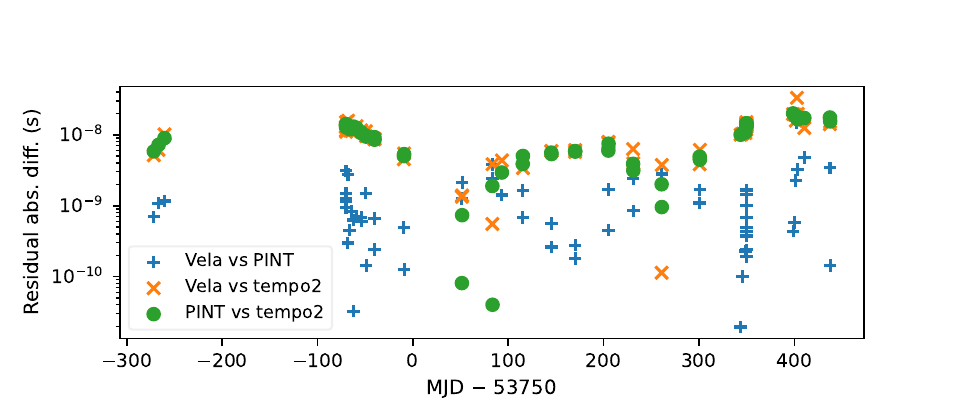}
    \caption{\updated{Absolute difference between the timing residuals computed using \vela{}, \pint{}, and \tempotwo{} for the same timing \& noise model given identical model parameters.  
    The dataset used herein corresponds to PSR J1748$-$2021E \citep{FreireRansom+2008} and is distributed as an example along with \pint{}.
    See subsection \ref{sec:example-NGC6440E} for more details.
    The three packages agree within $\sim10$ ns level.}}
    \label{fig:resdiff}
\end{figure}

\updated{The timing residuals computed using \vela{} agree with those computed using \pint{} and \tempotwo{} within the $\sim10$ ns level given identical timing \& noise models. 
An example comparison is shown in Figure \ref{fig:resdiff}.
The differences between timing residuals computed using \vela{}, \pint{}, and \tempotwo{} shown in this figure are slightly larger than, but of the same order of magnitude as, the difference between \pint{} and \tempotwo{} residuals shown in \citet{LuoRansom+2021}.
Further, we find better agreement between \vela{} and \pint{} as compared to either package with \tempotwo{}.
This is expected because \vela{} uses \pint{} for clock corrections and solar system ephemeris computations.
We suspect the difference between \vela{} and \pint{} residuals primarily arises due to the difference in the representation of extended precision numbers.}

\subsection{Quantities}
Dimensional analysis is an important tool for ensuring that the mathematical expressions describing the physical processes being studied are correctly translated into code.
It turns out that the pulsar timing formula given in equations \eqref{eq:delays}-\eqref{eq:resids} can be expressed such that all quantities therein have dimensions of the form $[\textsf{T}^d]$ where $d\in \mathbb{Z}$.
We give two examples below.
\begin{enumerate}
    \item The dispersion delay $\Delta_{\text{DM}}=\mathcal{K}\mathcal{D}\nu^{-2}$ where $\mathcal{K}$ is known as the dispersion constant, $\mathcal{D}$ is the dispersion measure (DM; the electron column density along the line of sight), and $\nu$ is the observing frequency in the SSB frame \citep{LorimerKramer2012}. 
    Although $\mathcal{D}$ has dimensions of $[\textsf{L}^{-2}]$, we can replace this parameter with the dispersion slope $\bar{\mathcal{D}} = \mathcal{K}\mathcal{D}$
    which has dimensions of $[\textsf{T}^{-1}]$.
    \item The binary Shapiro delay appearing in $\Delta_\text{B}$ for an eccentric orbit can be written as \citep{DamourDeruelle1986}
    $$\Delta_\text{BS}=\frac{-2GM_2}{c^3}\log\left[1-e\cos u - \sin\iota(\sin \omega (\cos u - e)+\sqrt{1-e^2}\cos\omega\sin u)\right]\,,$$
    where $M_2$ is the companion mass, $e$ is the eccentricity, $u$ is the eccentric anomaly, $\iota$ is the orbital inclination, and $\omega$ is the argument of periapsis.
    Here, the parameter $M_2$ has dimensions of $[\textsf{M}]$, but it can be replaced with $\bar{M}_2=GM_2/c^3$ which has dimensions of $[\textsf{T}]$.
\end{enumerate}

The above observation allows us to represent each quantity $q$ appearing in pulsar timing as a combination of a floating point number $x_q$ and a compile-time constant integer $d_q$, i.e., $q=x_q \, \text{s}^{d_q}$ where s represents the unit second, and the arithmetic of the $q$ objects follow the usual dimensional analysis rules.
Julia's Just-In-Time compilation allows arithmetic operations using this representation to be executed with almost zero run-time overhead, and this is implemented in the \gu{} package\footnote{Available at \url{https://github.com/abhisrkckl/GeometricUnits.jl/}} as the \texttt{GQ\{d,X<:AbstractFloat\}} type (`$<:$' represents `subtype of').
For example, a TOA value has the \texttt{GQ\{1,Double64\}} type and the observing frequency has the \texttt{GQ\{-1,Float64\}} type.
It also has the benefit of localizing the unit conversion operations to a certain part of the codebase, resulting in easier debugging.
It should be noted that time quantities like the TOA values and the various epochs appearing in the timing \& noise model are shifted such that the rotational frequency epoch ($t_0$ in equation \eqref{eq:phasing}) vanishes.

In comparison, \pint{} uses the \texttt{astropy.units} module \citep{RobitailleTollerud+2013} for this purpose which has a non-negligible computational overhead and \tempotwo{} does not enforce dimensional correctness in this way at all.

\subsection{TOAs}
A TOA value measured against an observatory clock is usually stored alongside the corresponding measurement uncertainty, observing frequency, a telescope code, and information about the observation setup represented as flags.
The various delay and phase corrections as well as the TOA covariance matrix appearing in equations \eqref{eq:delays}--\eqref{eq:lnlike} may in general depend on any subset of this information.

\begin{figure*}[t]
    \centering
    \includegraphics[width=0.7\linewidth]{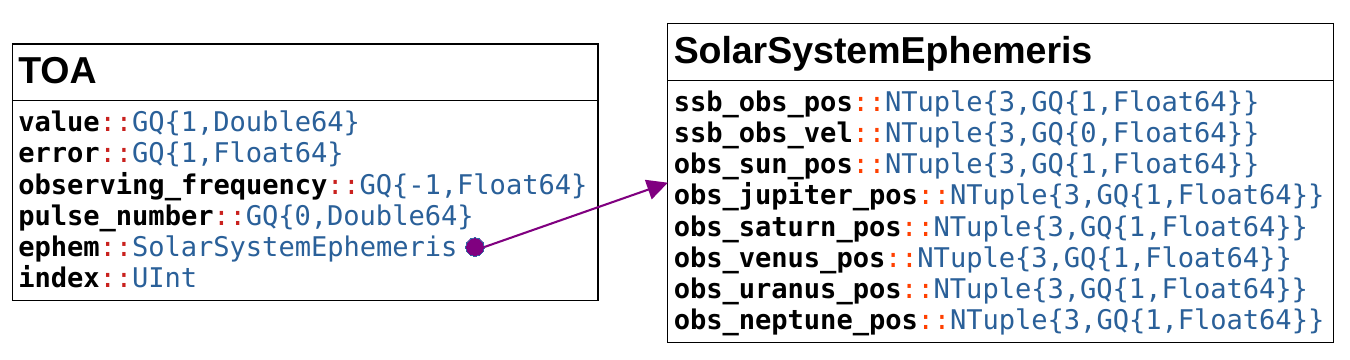}
    \caption{The structure of the \toa{} type. 
    \updated{`::' represents `is instance of'.}
    \texttt{TOA.value} is in the TDB timescale. The \texttt{SolarSystemEphemeris} type contains 3-tuples representing the position and velocity of the observatory with respect to the SSB (\texttt{SolarSystemEphemeris.ssb\_obs\_pos} and \texttt{SolarSystemEphemeris.ssb\_obs\_vel}), and the positions of various solar system objects with respect to the observatory.
    The clock corrections required for computing \texttt{TOA.value} and the solar system ephemerides are evaluated with the help of \pint{}.
    \texttt{UInt} is an unsigned integer.
    }
    \label{fig:toa_type}
\end{figure*}

A TOA is represented in \vela{} by the \toa{} type, whose structure is shown in Figure \ref{fig:toa_type}.
The \texttt{TOA.value} element represents the measured TOA value in \updated{Barycentric Dynamic Time \citep[TDB:][]{KlionerCapitaine+2009}}, \texttt{TOA.error} represents the measurement error, \texttt{TOA.observing\_frequency} represents the observing frequency in the observatory frame, \texttt{TOA.pulse\_number} is the pulse number corresponding to the TOA ($N[\phi]$ in equation \eqref{eq:resids}), \texttt{TOA.ephem} contains the solar system ephemerides evaluated at the TOA epoch, and \texttt{TOA.index} is an ordinal index.
The TOAs are usually stored and distributed as \tim{} files, and \vela{} uses \pint{} to read these files to create a collection of \toa{} objects.
The clock corrections needed for converting the TOA value from the observatory timescale to TDB and the solar system ephemerides are precomputed using \pint{}, which uses \astropy{} \citep{RobitailleTollerud+2013} internally; see subsection \ref{sec:pyvela}.
A dataset contains many TOAs, and this is represented as a \texttt{Vector\{TOA\}} object that stores the \toa{}s contiguously in memory.

This representation of the TOA is based on the following assumptions: (a) a phase-connected timing solution is available, and (b) the collection of TOAs is immutable\footnote{These assumptions are valid for single-pulsar noise \& timing analysis, which is performed after the data preparation and some preliminary timing is done. However, they do not apply to timing packages like \tempotwo{} or \pint{} because their use cases include adding/removing TOAs and editing TOA flags.}.
The first assumption allows us to pre-compute \texttt{TOA.pulse\_number}. 
The second assumption allows us to convert the TOA flags, which are string key-value pairs that are expensive to store and manipulate, into inexpensive bit masks specific to the timing model.
For example, if the timing model contains $N_{\text{jump}}$ observing system-dependent jumps ($\Delta_{\text{jump}}$) which depend on some TOA flags, the same information can be represented as an $N_{\text{toa}}\times N_{\text{jump}}$-dimensional bit mask (this is stored within the timing model instead of the TOAs).

\subsection{The timing \& noise model}
\label{sec:model-impl}
We begin this section by writing down a version of equation \eqref{eq:delays} that is closer to how it is implemented in practice. 
Some of the delays are omitted for simplicity.
\begin{equation}
    t_{\text{em}} + \mathcal{N} = {t}_{\text{tdb}} 
    - \Delta_{\odot}(\bar{t}_{\text{tdb}}) 
    - \Delta_{\text{DM}}(\bar{t}_{\text{ssb}})
    - \Delta_{\text{B}}(\bar{t}_{\text{dd}}) - ...\,,
\end{equation}
where 
$\bar{t}_X$ represents a TOA ($t_X$) along with its uncertainty ($\sigma_X$), observing frequency ($\nu_X$), etc after a correction step $X$,
$\mathcal{N}$ represents the uncorrelated noise present in the data, 
$t_{\text{tdb}}=t_{\text{arr}}-\Delta_{\text{clock}}(\bar{t}_{\text{arr}})$ is the TOA in the TDB timescale,
$t_{\text{ssb}}=t_{\text{tdb}}-\Delta_{\odot}(\bar{t}_{\text{tdb}})$ is the barycentered TOA, 
$\nu_{\text{ssb}}=\nu_{\text{tdb}}(1-\delta_{\odot}(\bar{t}_{\text{tdb}}))$ is the barycentered observing frequency,
$\delta_{\odot}$ is a Doppler factor due to solar system motion, and
$t_{\text{dd}}=t_{\text{ssb}}-\Delta_{\text{DM}}(\bar{t}_{\text{ssb}})$ 
is the dedispersed TOA.
The above expression makes it clear that the delay corrections must be applied in a specific order: solar system delays, interstellar dispersion delays, pulsar binary delays, etc.
\textit{Then}, the different phase terms in equation \eqref{eq:phasing} are computed.

In \vela{}, the different delay and phase terms in equations \eqref{eq:delays} and \eqref{eq:phasing} and the TOA uncertainty corrections in equation \eqref{eq:errcorr} that \updated{can be computed independently for each TOA} are implemented as subtypes of the \comp{} type. 
The hierarchy and descriptions of various \comp{} types available in \vela{} are given in Appendix \ref{sec:component-types}.
Each \comp{} type has an associated \texttt{correct\_toa()} method which computes the corresponding delay, phase, Doppler factor, uncertainty correction, etc, \updated{as applicable}.
The corrected TOA $t_{\text{em}}$, the phase $\phi(t_{\text{em}})$, the scaled TOA uncertainty $\varsigma$, and the topocentric frequency $F$ are computed by successively applying these methods in the correct order.
Finally, the timing residuals are computed using equation \eqref{eq:resids}.

The likelihood function is computed from the timing residuals with the help of a \kernel{} object which represents the matrix operations present in equation \ref{eq:lnlike}.
Two \kernel{} types are available currently.
\texttt{WhiteNoiseKernel} represents the case when only uncorrelated (white) noise is present in the TOAs, i.e., the covariance matrix $\textbf{C}=\textbf{N}$ is diagonal.
In this case, $\ln L$ can be evaluated with $\mathcal{O}(N_{\text{toa}})$ time complexity.
\texttt{EcorrKernel} represents the case when only time-uncorrelated noise (i.e., white noise and ECORR) is present.
It turns out that $\ln L$ can be evaluated with $\mathcal{O}(N_{\text{toa}})$ time complexity in the latter case also, see Appendix \ref{sec:ecorr} for details. 

\begin{figure*}[t]
    \centering
   \includegraphics[width=0.4\linewidth]{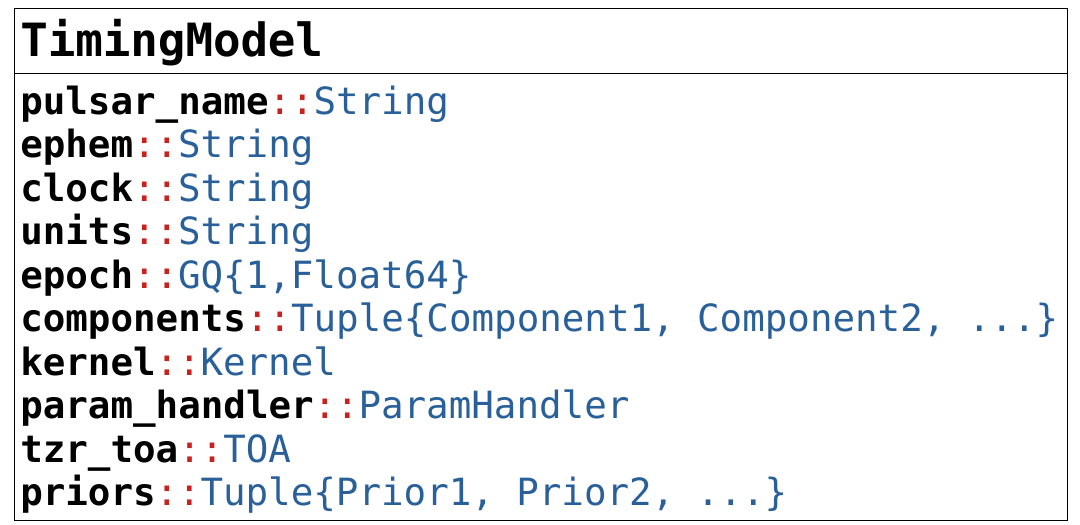}
    \caption{The structure of the \tm{} type.
    \updated{`::' represents `is instance of'.}
    \texttt{pulsar\_name} stores the name of the pulsar. 
    \texttt{ephem} is the name of the solar system ephemeris model.
    \texttt{clock} is the name of the realization of the TT timescale used in clock corrections (e.g., TT(BIPM2021)); see \citet{HobbsEdwardsManchester2006}.
    \texttt{units} is the timescale of \texttt{TOA.value} (only TDB is supported).
    \texttt{epoch} is the rotational frequency epoch.
    The \texttt{components} tuple contains \comp{} objects representing different terms in the timing \& noise model that are uncorrelated across TOAs.
    \kernel{} is an object representing the computation in equation \eqref{eq:lnlike} including the various contributions to the TOA covariance matrix \textbf{C}.
    \texttt{param\_handler} contains information about the different parameters appearing in the timing \& noise model.
    \texttt{tzr\_toa} is a fiducial TOA with respect to which the rotational phase is measured (`tzr' stands for `t-zero').
    \texttt{priors} is a tuple that contains the prior distributions for the free model parameters.
    }
    \label{fig:model_type}
\end{figure*}

The timing \& noise model is represented in \vela{} as the \tm{} type, shown in Figure \ref{fig:model_type}.
In addition to the \comp{}s and the \kernel{} described above, a \tm{} object also contains a \texttt{ParamHandler} object that contains information about the model parameters and an ordered collection of \texttt{Prior} objects which implement the prior distributions appearing in equation \ref{eq:lnpost}; see Appendix \ref{sec:priors} for more details.

\vela{} provides functions that return callable objects which compute the log-likelihood (\texttt{get\_lnlike\_func()}), the log-prior (\texttt{get\_lnprior\_func()}), the prior transform (\texttt{get\_prior\_transform\_func()}), and the log-posterior (\texttt{get\_lnpost\_func()}).
The outputs of these functions can be passed on to any MCMC or nested sampler.

Three parallelization paradigms are provided for the log-likelihood and log-posterior computation.
By default, the computation is parallelized across TOAs using multi-threading (equations \eqref{eq:lnlike} and \eqref{eq:ln_Lab} are trivially parallelizable this way).
Some ensemble samplers like \emcee{} support vectorized execution of the posterior distribution across multiple points in the parameter space, and \vela{} provides the option to do this parallelly using multiple threads.
Alternatively, \vela{} also provides serial versions of the same functions for cases where the sampler itself implements parallelization, e.g., using Message Passing Interface \citep[MPI:][]{GroppLusk+1996}.

\subsection{The \pyvela{} Interface}
\label{sec:pyvela}
We provide a Python interface for \vela{} called \pyvela{} developed using \texttt{JuliaCall/PythonCall} \citep{Rowley2022} for easy usage. 
This is useful because Pulsar Astronomers tend to be more familiar with Python than Julia, and because Python offers a wider choice of sampling packages than Julia.

The \spnta{} class (SPNTA stands for single-pulsar noise \& timing analysis) in \pyvela{} reads a pair of \parf{} and \tim{} files, performs the clock corrections and solar system ephemeris computations with the help of \pint{}, constructs the prior distributions (see Appendix \ref{sec:priors}), and creates a pair of \tm{} and \texttt{Vector\{TOA\}} objects described above.
It also provides a convenient interface for evaluating the log-likelihood, log-prior, prior transform, and log-posterior functions.
An example code snippet using \pyvela{} with the MCMC sampler \emcee{} \citep{Foreman-MackeyHogg2013} is shown in Figure \ref{algo:pyvela-emcee}, and an example using \pyvela{} with the nested sampler \texttt{nestle} \citep{Barbary2021} is shown in Figure \ref{algo:pyvela-nestle}.

\begin{figure*}[p]
\begin{lstlisting}[language=Python]
from pyvela import SPNTA
import emcee
import numpy as np

spnta = SPNTA(
    parfile="NGC6440E.par",
    timfile="NGC6440E.tim", 
    custom_priors="NGC6440E_priors.json",
)

nwalkers = spnta.ndim * 5
p0 = np.array([
    spnta.prior_transform(cube) 
    for cube in np.random.rand(nwalkers, spnta.ndim)
])

sampler = emcee.EnsembleSampler(
    nwalkers,
    spnta.ndim,
    spnta.lnpost,
)

sampler.run_mcmc(p0, 6000, progress=True)
samples_raw = sampler.get_chain(flat=True, discard=1000, thin=50)

samples = spnta.rescale_samples(samples_raw)
\end{lstlisting}
\caption{An example code snippet demonstrating the \pyvela{} interface with the MCMC sampler \emcee{}. 
User-defined priors are read from a \texttt{JSON} file (see Appendix \ref{sec:priors}).
The \texttt{emcee.EnsembleSampler} object is initialized with samples drawn from the prior distribution with the help of \texttt{SPNTA.prior\_transform()}. The samples are originally in \vela's internal units. They are converted into their commonly used units using \texttt{SPNTA.rescale\_samples()}.}
\label{algo:pyvela-emcee}
\end{figure*}

\begin{figure*}[p]
\begin{lstlisting}[language=Python]
from pyvela import SPNTA
import nestle

spnta = SPNTA(
    parfile="NGC6440E.par",
    timfile="NGC6440E.tim", 
    custom_priors="NGC6440E_priors.json",
)

results = nestle.sample(
    spnta.lnlike,
    spnta.prior_transform,
    spnta.ndim,
    method="multi",
    npoints=500,
    dlogz=0.001,
    callback=nestle.print_progress,
)

samples_raw = nestle.resample_equal(results.samples, results.weights)
samples = spnta.rescale_samples(samples_raw)

log_evidence = results.logz
\end{lstlisting}
\caption{An example code snippet demonstrating the \pyvela{} interface with the nested sampler \texttt{nestle}.
\texttt{method="multi"} invokes the \texttt{MultiNest} algorithm \citep{FerozHobson+2008}.
}
\label{algo:pyvela-nestle}
\end{figure*}

\begin{figure*}[t]
    \centering
    \includegraphics[width=1\linewidth]{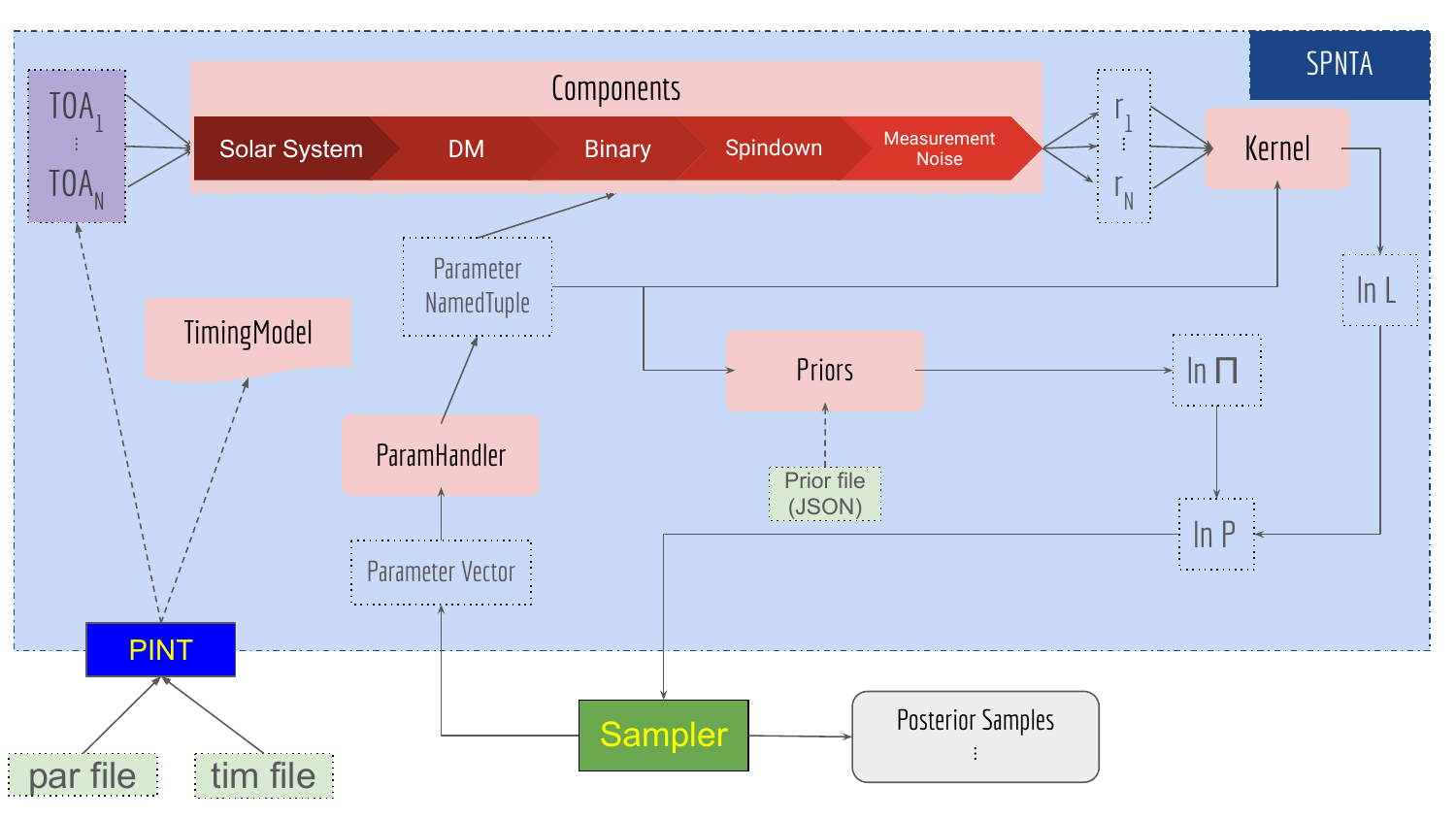}
    \caption{A schematic diagram summarizing \vela{}'s architecture. \comp{}s, \kernel{}, \texttt{Prior}s, and \texttt{ParamHandler} are parts of the \tm{}.}
    \label{fig:schematic}
\end{figure*}

A schematic diagram that summarizes how \vela{} works is shown in Figure \ref{fig:schematic}.

\section{Application to simulated datasets}
\label{sec:examples}
In this section, we demonstrate the application of \vela{} on two simulated datasets.
These are generated using the \texttt{pint.simulation} module \citep{SusobhananKaplan+2024} based on certain real datasets.

\subsection{PSR J1748\textminus{}2021E (NGC 6440E)}
\label{sec:example-NGC6440E}
PSR J1748-2021E is an isolated pulsar located in the globular cluster NGC 6440 \citep{FreireRansom+2008}.
\updated{The simulated dataset is generated based on a real dataset distributed as an example alongside \pint{} \citep{LuoRansom+2021}} and contains 61 TOAs taken using the Green Bank Telescope from 2005 to 2006 with observing frequencies in the range 1550-1212 MHz.

We convert the TOAs from the observatory timescale to TDB with the help of the BIPM2021 realization of the TT timescale and the DE421 solar system ephemeris.
The timing \& noise model includes solar system delays, interstellar dispersion, spindown, \updated{an overall phase offset,} and a global EFAC that scales the measured TOA uncertainties.
This model has 7 free parameters.

We run the Bayesian analysis using \vela{} with \emcee{}, which implements the affine-invariant ensemble sampler algorithm \citep{Foreman-MackeyHogg2013} (see Figure \ref{algo:pyvela-emcee} for the Python script)\footnote{This MCMC run was executed for 6000 steps with 35 walkers on an AMD Ryzen 7 CPU with 16 logical cores using 4 threads. It took approximately 2 seconds.}.
The prior distribution of the global EFAC is LogNormal[0,0.25] and that of the overall phase offset is Uniform[-0.5,0.5].
`Cheat' prior distributions, i.e., uniform distributions centered at the maximum likelihood values obtained using \pint{} whose widths are 10 times the frequentist uncertainties, are used for all other parameters (see Appendix \ref{sec:priors}).
We have checked that increasing the width of the `cheat' priors does not appreciably alter the posterior distribution.

\begin{figure*}[p]
    \centering
    \includegraphics[width=1\linewidth]{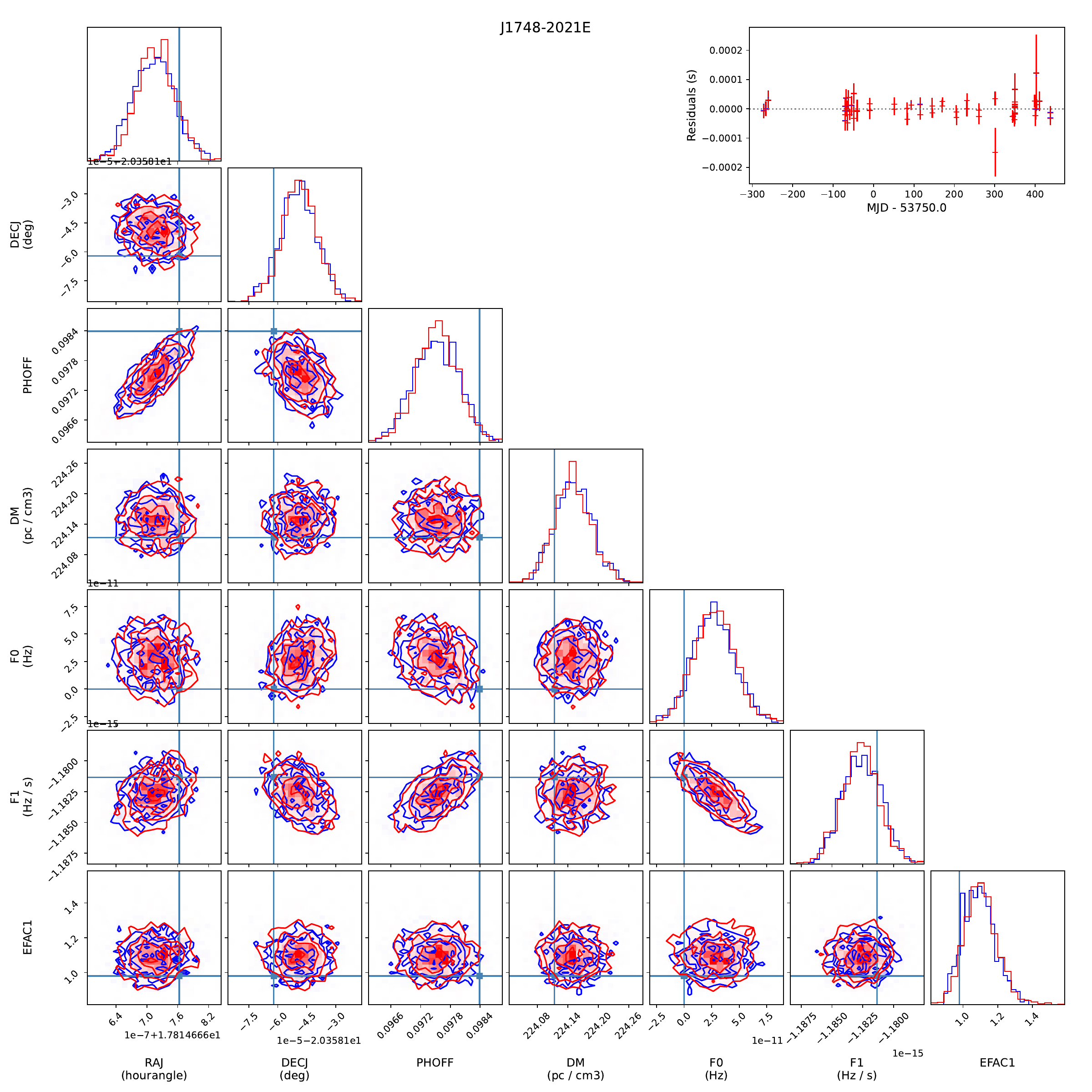}
    \caption{Bayesian timing \& noise analysis results for a simulated dataset of PSR J1748$-$2021E. 
    The plotted parameters are source coordinates (RAJ, DECJ), overall phase offset (PHOFF), dispersion measure (DM), rotational phase and its derivative (F0, F1), and a global EFAC (EFAC1).
    The results obtained using \vela{} are plotted in red, and those obtained using \pint{} are plotted in blue. The inset shows the timing residuals computed using the median values of the posterior samples. A value of 61.4854765543713046 Hz has been subtracted from the rotational frequency F0.} 
    \label{fig:NGC6440E}
\end{figure*}

We repeat this analysis using the \texttt{pint.bayesian} module \citep{SusobhananKaplan+2024} with \emcee{} for comparison. 
This is computationally feasible due to the small size of the dataset.
The posterior distributions and post-fit residuals obtained from this exercise are shown in Figure \ref{fig:NGC6440E}, and show good agreement between the posterior distributions and post-fit residuals obtained using \vela{} and \texttt{pint.bayesian}.
Further, the estimated parameters agree with the injected parameters within 2$\sigma$ uncertainties.

\subsection{PSR J1909\textminus{}3744}
PSR J1909$-$3744 is a millisecond binary pulsar observed as part of multiple PTA campaigns. 
The simulated dataset used in this section is generated based on a subset of the narrowband data of  J1909$-$3744 published as part of the Indian Pulsar Timing Array data release 1 \citep[InPTA DR1:][]{TarafdarNobleson+2022}.
It contains 361 TOAs measured using the Giant Metrewave Radio Telescope (GMRT) during 2020-2021 in two frequency bands (300-500 MHz and 1260-1460 MHz).
Notably, we have injected the dispersion measure variations based on the epoch-wise DM measurements given in the InPTA DR1.

We convert the TOAs from the observatory timescale to TDB with the help of the BIPM2019 realization of the TT timescale and the DE440 solar system ephemeris.
The timing \& noise model used to fit the simulated data includes solar system delays (including parallax and proper motion), dispersion measure variations modeled as a combination of a Taylor series (up to the second DM derivative) and a Fourier series Gaussian process with 40 harmonics (see Appendix \ref{sec:rednoise}), the ELL1 model for a nearly circular binary, pulsar spindown, and EFACs and EQUADs applied to the two observing frequency bands.
This model has 104 free parameters in total.

The prior distributions of some of the parameters are given in Table \ref{tab:1909-priors}. 
The priors for the scaled Fourier amplitudes of the DM noise are unit normal distributions as discussed in Appendix \ref{sec:rednoise}.
`Cheat' priors are used for the other parameters with width 40 times their frequentist uncertainty.
We have checked that increasing the width of the `cheat' priors does not appreciably alter the posterior distribution.

\begin{table*}[tp]
\begin{tabular}{cll}
\hline
Parameter & Description  &  Prior \\\hline\hline
 M2    & Companion mass ($M_{\odot}$) & Uniform distribution around the IPTA DR2 measurement  \\
 & & with width 40 times the corresponding uncertainty \\\hline 
 SINI & Orbital inclination & $\Pi(\sin\iota) = \frac{\sin\iota}{\sqrt{1-\sin^2\iota}}$ (see Appendix \ref{sec:priors})\\\hline
 TNDMAMP & Spectral log-amplitude of DM noise & 
 LogUniform[$10^{-18}$, $10^{-12}$]\\\hline
 TNDMGAM & Spectral index of DM noise & Uniform[0.5, 7]\\\hline
 PHOFF & Overall phase offset & Uniform[-0.5, 0.5] \\\hline
 EFAC & Scale factor for TOA uncertainties & LogNormal[0, 0.25]\\\hline
 EQUAD & TOA uncertainty correction  & LogUniform[$10^{-3}$, $10$]\\
 & added in quadrature ($\mu$s) &\\\hline
\end{tabular}
\caption{The prior distributions used for analyzing the simulated dataset of PSR J1909$-$3744. `Cheat' priors are used for timing parameters not mentioned here. IPTA DR2 is the International Pulsar Timing Array Data Release 2 \citep{PereraDeCesar+2019}.}
\label{tab:1909-priors}
\end{table*}

\begin{figure*}[p]
    \centering
    \includegraphics[width=1\linewidth]{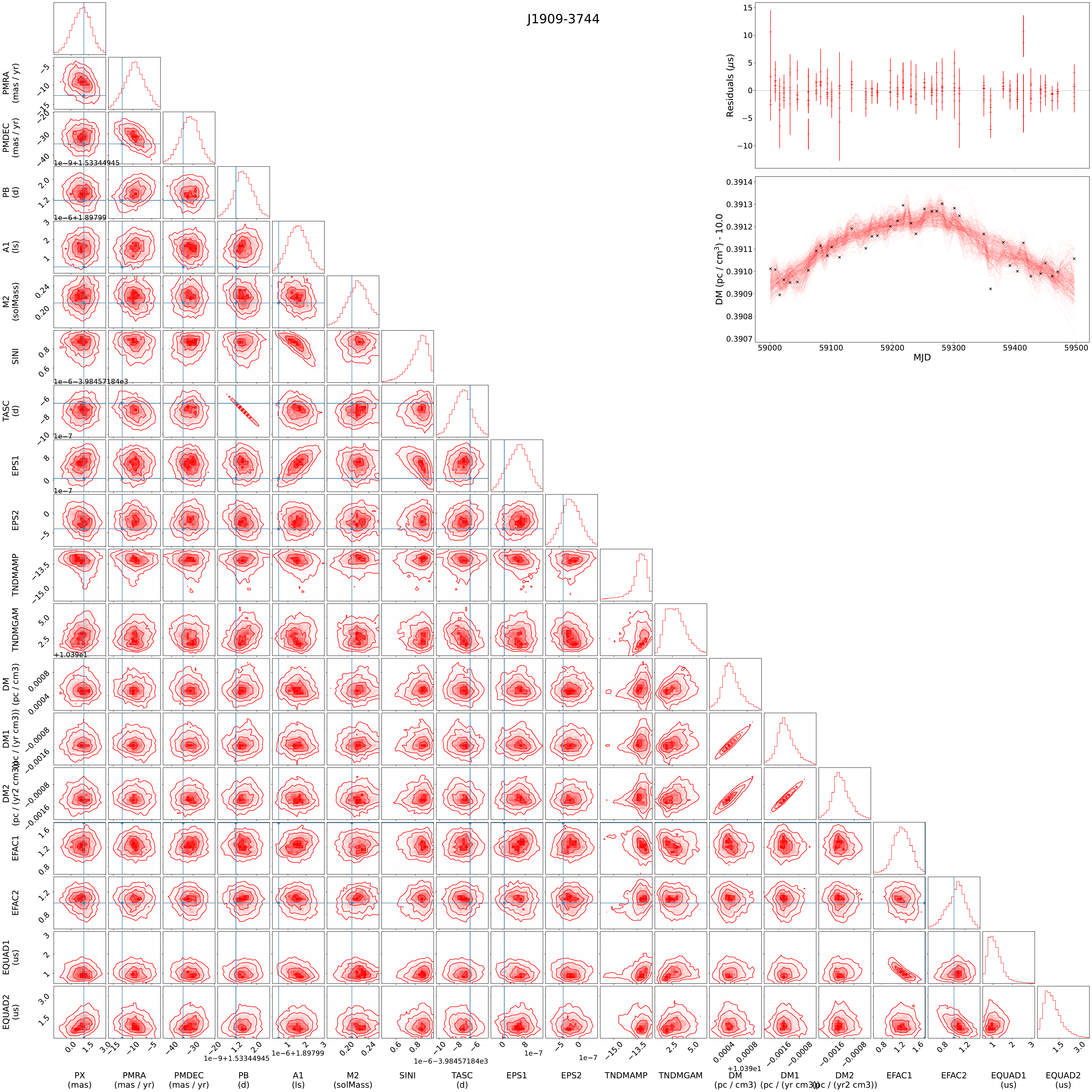}
    \caption{Bayesian timing \& noise analysis results for a simulated dataset of PSR J1909-3744.
    The plotted parameters are parallax (PX), proper motion (PMRA, PMDEC), orbital period (PB), projected semi-major axis of the pulsar orbit (A1), companion mass (M2), sine-inclination (SINI), epoch of ascending node (TASC), Laplace-Lagrange parameters (EPS1, EPS2), dispersion measure and its derivatives (DM, DM1, DM2), power law spectral parameters of the DM noise (TNDMAMP, TNDMGAM),  EFACs, and EQUADs. 
    The two EFACs and EQUADs correspond to the 300-500 MHz band and the 1260-1460 MHz band in that order.
    The source coordinates (RAJ, DECJ), the rotational frequency and its derivative (F0, F1),  the overall phase offset (PHOFF), and the Fourier amplitudes of the DM noise are not plotted.
    The vertical and horizontal lines represent the injected values.
    The estimated parameters of astrometric and binary parameters are consistent with the injected values within $3\sigma$ uncertainties.
    Note that a Fourier series Gaussian process model was used for the analysis whereas the injected DM values were taken from real epoch-wise measurements.
    The inset shows post-fit residuals in the top panel.
    The bottom panel of the inset shows the injected DM time series (black points) and 200 random realizations of the DM time series drawn from the posterior distribution.
}
\label{fig:J1909-3744}
\end{figure*}

The posterior samples were obtained using \emcee{}\footnote{Executed for 6000 steps with 520 walkers on an AMD Ryzen 7 CPU with 16 logical cores with 16 threads. It took approximately 4.5 minutes.}, and 
the results of this analysis are plotted in Figure \ref{fig:J1909-3744}.
We see that some of the estimated parameters do not agree well with their injected values.
This may be because the DM model is not adequately modeling some of the short-timescale DM variations (see inset of Figure \ref{fig:J1909-3744}).

\section{Summary \& Future directions}
\label{sec:summary}

We have developed a new package for performing Bayesian single-pulsar timing \& noise analysis named \vela{}.
This package is written in Julia and provides an efficient and parallelized implementation of the non-linear timing \& noise model along with a Python interface named \pyvela{}.
It uses \pint{} to read \parf{} \& \tim{} files, apply clock corrections to the TOAs, and compute solar system ephemerides but provides an independent implementation of the timing \& noise model.
Given a set of TOAs and a timing \& noise model, \vela{} provides an intuitive interface for computing the log-likelihood, log-prior, prior transform, and log-posterior functions that are compatible with various MCMC and nested samplers available in Julia and Python. 
The architecture of \vela{} is summarized below.
\begin{enumerate}
    \item Dimensionful quantities are converted to a form that has dimensions $[\textsf{T}^d]$ and represented using \texttt{GQ\{d,X<:AbstractFloat\}} types. Values that require extended precision are stored using the \texttt{Double64} type.
    \item Clock-corrected TOAs and their metadata along with the pre-computed solar system ephemerides are stored using the \toa{} type.
    \item The timing \& noise model is represented using the \tm{} type. It contains:
    \begin{enumerate}
        \item \comp{}s which represent various astrophysical and instrumental effects that are uncorrelated across TOAs.
        \item A \kernel{} which represents the log-likelihood computation and correlated noise if any.
        \item \texttt{Prior}s which represent the prior distributions of free model parameters.
        \item A \texttt{ParamHandler} which converts the parameter values provided by the sampler into a representation that can be accessed efficiently.
    \end{enumerate}
    \item The \pyvela{} interface provides a Python binding to \vela{}. It contains the \spnta{} class which
    \begin{enumerate}
        \item Reads the \parf{} and \tim{} files, computes clock corrections and solar system ephemerides, and constructs the \tm{} and \toa{} objects.
        \item Provides methods that compute the log-likelihood, log-prior, prior transform, and log-posterior that can be passed into samplers.
    \end{enumerate}
\end{enumerate}

We demonstrated the usage of \vela{} using two example datasets.
In the case of the smaller dataset, we showed that the parameter estimation results are consistent with those estimated using \pint{}.

\vela{} was developed to be a more user-friendly and flexible alternative to \temponest{}, and it should complement the capabilities of \enterprise{} which does not allow the exploration of the full non-linear timing model.
\vela{} will be a useful tool for analyzing the various pulsar timing datasets, especially those of high-precision experiments such as PTAs that are rapidly growing in volume and sensitivity.

The planned future development wishlist for \vela{} is as follows.
\begin{enumerate}
    \item Develop a vibrant community of users and developers.
    \item Implement wideband timing \citep{AlamArzoumanian+2021} (under development).
    \item Implement photon domain timing for high-energy pulsar timing \citep{PletschClark+2015} and enable simultaneous analysis of radio TOA data and high-energy photon data.
    \item Implement a sampler that is tuned to the pulsar timing \& noise analysis. Possible avenues to explore are Gibbs sampling \citep[e.g.][]{LaalLamb+2023} and Hamiltonian Monte Carlo \citep[e.g.][]{FreedmanJohnson+2023}.
    \item Implement automatic differentiation using a tool like \texttt{Zygote.jl} \citep{Innes2018}. This is crucial for implementing Hamiltonian Monte Carlo.
    \item Implement a fitter for \pint{} using \vela{} as a backend. This should be useful in cases where \pint{} is too slow, especially for noise estimation \citep{SusobhananKaplan+2024}.
    \item Implement timing model components that are currently unavailable in \vela{}, such as the DDGR binary model \citep{DamourDeruelle1986}, advanced solar wind models \citep{HazbounSimon+2022,SusarlaChalumeau+2024}, etc.
\end{enumerate}

\section*{Data Availability}
The dataset of PSR J1748$-$2021E used to generate the simulated dataset is distributed with \pint{} as an example dataset at \url{https://github.com/nanograv/PINT/}.
The dataset of PSR J1909$-$3744 used to generate the simulated dataset is available as part of the InPTA DR1 at \url{https://github.com/inpta/InPTA.DR1/}.

\section*{Acknowledgments}
AS thanks David Kaplan for valuable suggestions on the manuscript and Tjonnie Li and Rutger van Haasteren for fruitful discussions.
AS thanks the anonymous referee for valuable comments and suggestions.

%



~

\software{
\pint{} \citep{LuoRansom+2021, SusobhananKaplan+2024}, 
\emcee{} \citep{Foreman-MackeyHogg2013}, 
\texttt{nestle} \citep{Barbary2021},
\texttt{numpy} \citep{HarrisMillman+2020}, 
\astropy{} \citep{RobitailleTollerud+2013}, 
\texttt{matplotlib} \citep{Hunter2007}, 
\texttt{corner} \citep{Foreman-Mackey2016}, 
\texttt{DoubleFloats.jl} \citep{Sarnoff2022}, \texttt{Distributions.jl} \citep{BesançonPapamarkou+2021}, \texttt{JuliaCall/PythonCall.jl} \citep{Rowley2022},
\texttt{git}\footnote{\url{https://git-scm.com/}} 
}



\appendix

\section{Component types available in \vela{}}
\begin{figure*}[htp]
    \centering
    \includegraphics[width=1.05\linewidth]{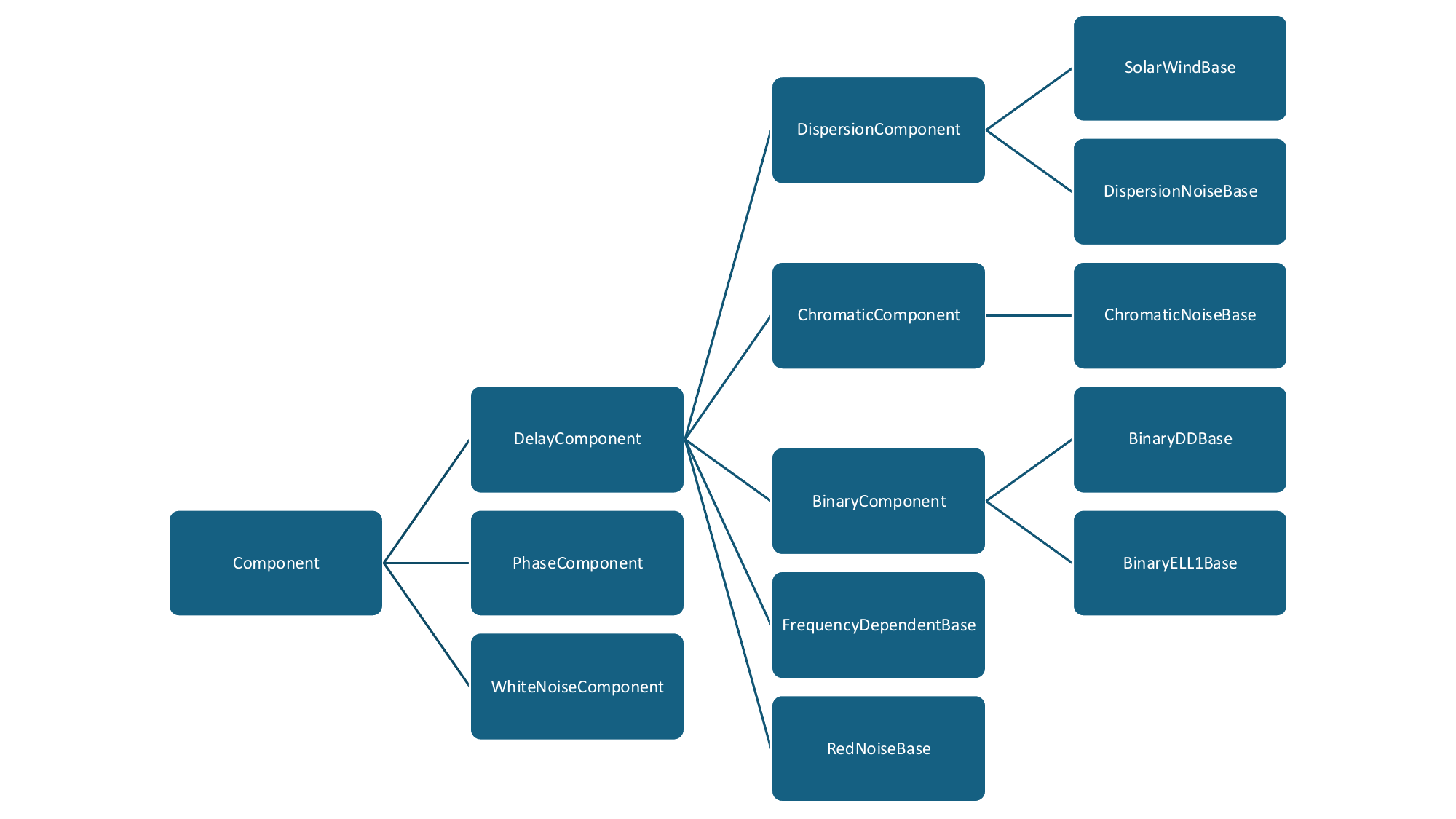}
    \caption{The hierarchy of \comp{} base types. See Table \ref{tab:components} for their implemented concrete subtypes.}
    \label{fig:component-base-types}
\end{figure*}

Figure \ref{fig:component-base-types} shows the hierarchy of the various abstract subtypes of \comp{}.
Concrete \comp{} types are listed in Table \ref{tab:components}.

\label{sec:component-types}

\begin{longtable*}{lll}
\hline
\textbf{Component \& Base Type}& \textbf{\pint{} Equivalent(s)}             & \textbf{Description \& Reference}                                                \\ \hline\hline
\texttt{SolarSystem}                         & \texttt{AstrometryEquatorial}                 & Solar system delays (R{\o}mer, parallax, and  Shapiro) and Doppler                                    \\
\texttt{\textless{}: DelayComponent}         & \texttt{AstrometryEcliptic}                   &  correction  \citep{EdwardsHobbsManchester2006}                                             \\
                                    & \texttt{SolarSystemShapiro}                   &                                                                   \\ \hline
\texttt{SolarWindDispersion}                 & \texttt{SolarWindDispersion} & Solar wind dispersion assuming spherical   symmetry                                     \\
\texttt{\textless{}: SolarWindBase}          &                                      &   \citep{EdwardsHobbsManchester2006}                                                    \\\hline
\texttt{DispersionTaylor}                    & \texttt{DispersionDM}                         & Taylor series representation of interstellar  dispersion                                   \\
\texttt{\textless{}: DispersionComponent}    &                                      &  \citep{BackerHellings1986}                                             \\\hline
\texttt{DispersionPiecewise}                    & \texttt{DispersionDMX}                         & Piecewise-constant representation of interstellar  dispersion                                   \\
\texttt{\textless{}: DispersionComponent}    &                                      &  \citep{ArzoumanianBrazier+2015}                                             \\\hline
\texttt{DMWaveX}                             & \texttt{DMWaveX}                              & Unconstrained Fourier series representation  of interstellar                     \\
\texttt{\textless{}: DispersionNoiseBase}    &                                      &  dispersion variations  \citep{SusobhananKaplan+2024}                                          \\ \hline
\texttt{PowerlawDispersionNoiseGP}           & \texttt{PLDMNoise}                            & Fourier series Gaussian process representation  of interstellar                                  \\
\texttt{\textless{}: DispersionNoiseBase}    &                                      &  dispersion variations with a power law spectrum \citep{LentatiHobson+2014}                              \\\hline
\texttt{DispersionOffset}                    & \texttt{FDJumpDM}                             & System-dependent dispersion measure offset                                       \\
\texttt{\textless{}: DispersionComponent}    &                                      &                                                                                  \\\hline
\texttt{DispersionJump}                    & \texttt{DispersionJump}                             & System-dependent dispersion measure offset without a delay (for                                       \\
\texttt{\textless{}: DispersionComponent}    &                                      &                                                                   wideband  data)   \citep{AlamArzoumanian+2021}           \\\hline
\texttt{ChromaticTaylor}                     & \texttt{ChromaticCM}                          & Taylor series representation of a chromatic  delay                                  \\
\texttt{\textless{}: ChromaticComponent}     &                                      &                                                     \\\hline
\texttt{ChromaticPiecewise}                     & \texttt{ChromaticCMX}                          & Piecewise-constant representation of a chromatic  delay                                  \\
\texttt{\textless{}: ChromaticComponent}     &                                      &   \citep{HembergerStinebring2008}                                                  \\\hline
\texttt{CMWaveX}                             & \texttt{CMWaveX}                              & Unconstrained Fourier series representation of variable chromatic                                  \\
\texttt{\textless{}: ChromaticNoiseBase}     &                                      &                                                       delay  \\\hline
\texttt{PowerlawChromaticNoiseGP}            & \texttt{PLChromNoise}                         & Fourier series Gaussian  process representation  of variable chromatic                                \\
\texttt{\textless{}: ChromaticNoiseBase }    &                                      &  delay with a power law spectrum \citep{LentatiHobson+2014}                                    \\\hline
\texttt{BinaryDD}                            & \texttt{BinaryDD}                             & Parameterized model for eccentric binaries   with relativistic effects                                    \\
\texttt{\textless{}: BinaryDDBase}           & \texttt{BinaryBT}                             &  \citep{DamourDeruelle1986}                                     \\\hline
\texttt{BinaryDDH}                           & \texttt{BinaryDDH}                            & Similar to \texttt{BinaryDD}, but with an orthometric  representation of Shapiro                                       \\
\texttt{\textless{}: BinaryDDBase}           &                                      &     delay suitable for low-inclination systems \citep{WeisbergHuang2016}                                \\ \hline
\texttt{BinaryDDK}                           & \texttt{BinaryDDK}                            & Similar to \texttt{BinaryDD}, but with Kopeikin corrections due to parallax                               \\
\texttt{\textless{}: BinaryDDBase}           &                                      & and proper motion \citep{Kopeikin1995,Kopeikin1996}                          \\\hline
\texttt{BinaryDDS}                           & \texttt{BinaryDDS}                            & Similar to \texttt{BinaryDD}, but with an alternative parameterization of                                    \\
\texttt{\textless{}: BinaryDDBase}           &                                      &  Shapiro delay suitable for   almost edge-on systems \citep{RafikovLai2006}                                 \\\hline
\texttt{BinaryELL1}                          & \texttt{BinaryELL1}                           & Parameterized model for almost circular binaries with     relativistic                       \\
\texttt{\textless{}: BinaryELL1Base}         &                                      &  effects \citep{LangeCamilo+2001}                                         \\\hline
\texttt{BinaryELL1H}                         & \texttt{BinaryELL1H}                          & Similar to \texttt{BinaryELL1}, but with an orthometric   representation of                                   \\
\texttt{\textless{}: BinaryELL1Base}         &                                      &  Shapiro delay \citep{FreireWex2010}                               \\\hline
\texttt{BinaryELL1k}                         & \texttt{BinaryELL1k}                          & Similar to \texttt{BinaryELL1}, but with an exact treatment of advance of                       \\
\texttt{\textless{}: BinaryELL1Base}         &                                      &    periapsis \citep{SusobhananGopakumar+2018}                               \\\hline
\texttt{FrequencyDependent}                  & \texttt{FD}                                   & Phenomenological model for apparent delays caused by un-modeled           \\
\texttt{\textless{}: FrequencyDependentBase} &                                      & profile evolution  \citep{ArzoumanianBrazier+2015}                                                                      \\\hline
\texttt{FrequencyDependentJump}              & \texttt{FDJump}                               & Similar to \texttt{FrequencyDependent}, but accounts for experiment-dependent  \\
\texttt{\textless{}: FrequencyDependentBase} &                                      & differences in modeling profile evolution \citep{SusobhananKaplan+2024}                                                  \\\hline
\texttt{WaveX}                               & \texttt{WaveX}                                & Apparent delay due to rotational irregularities of the pulsar  \\
\texttt{\textless{}: RedNoiseBase}           &                                      & represented as an  unconstrained Fourier series  \citep{SusobhananKaplan+2024}                                                   \\\hline
\texttt{PowerlawRedNoiseGP}                  & \texttt{PLRedNoise}                           & Rotational irregularities of the pulsar represented as a Fourier   \\
\texttt{\textless{}: RedNoiseBase}           &                                      &  series Gaussian process with power law spectrum  \citep{LentatiHobson+2014}                        \\\hline
\texttt{Spindown}                            & \texttt{Spindown}                             & Taylor series representation of the pulsar spin-down \citep{HobbsEdwardsManchester2006}                             \\
\texttt{\textless{}: PhaseComponent}         &                                      &                                                                                  \\\hline
\texttt{Glitch}                              & \texttt{Glitch}                               & Pulsar glitches  \citep{HobbsEdwardsManchester2006}                                                               \\
\texttt{\textless{}: PhaseComponent}         &                                      &                                                                                  \\\hline
\texttt{PhaseOffset}                         & \texttt{PhaseOffset}                          & Overall initial phase  \citep{SusobhananKaplan+2024}                                                            \\
\texttt{\textless{}: PhaseComponent}         &                                      &                                                                                  \\\hline
\texttt{PhaseJump}                           & \texttt{PhaseJump}                            & System-dependent phase offsets      \citep{HobbsEdwardsManchester2006}                                             \\
\texttt{\textless{}: PhaseComponent}         &                                      &                                                                                  \\\hline
\texttt{MeasurementNoise}                    & \texttt{ScaleToaErrors}                       & Corrections to the measured TOA uncertainties  \citep{LentatiHobson+2014}                                  \\
\texttt{\textless{}: WhiteNoiseComponent}&                                      &     \\\hline   
\texttt{DispersionMeasurementNoise}                    & \texttt{ScaleDmErrors}                       & Corrections to the measured wideband DM uncertainties                                    \\
\texttt{\textless{}: WhiteNoiseComponent}&                                      &  \citep{AlamArzoumanian+2021}   \\\hline
\caption{Timing \& noise model components available in \vela{}. See Section \ref{sec:timing-basics} for an overview of the timing \& noise model and Section \ref{sec:model-impl} for its implementation in \vela{}. See Figure \ref{fig:component-base-types} for the hierarchy of base types.  The `\texttt{\textless{}:}' symbol represents `subtype of'.}
\label{tab:components}
\end{longtable*}

\section{Prior distributions}
\label{sec:priors}

In principle, the prior for each parameter should be set strictly based on our prior knowledge. 
Indeed, we may have prior information on some of the parameters from previous timing experiments, VLBI campaigns \citep[e.g.][]{DellerVigeland+2016}, detection of counterparts in other parts of the electromagnetic spectrum \citep[e.g.][]{KaplanStovall+2018}, etc. 
Or priors may be estimated from population statistics using a catalog like \texttt{psrcat} \citep{ManchesterHobbs+2005}.

However, for many parameters, pulsar timing provides so much signal-to-noise ratio (S/N) that the effect of the prior on the posterior distribution is entirely negligible provided the prior is sufficiently broad. 
This is the case for parameters like $F_0$, $F_1$, source coordinates, etc even for small timing datasets.
In the context of analytic marginalization of linearized timing model parameters, it is customary to assume uninformative infinitely broad Gaussian priors \citep{van_HaasterenLevin2013}.
On the other hand, given the insensitivity of the posterior distribution on the priors on some of the timing model parameters, `cheat' priors, namely uniform distributions centered around the frequentist estimate whose width is several times (e.g., 10x) the frequentist uncertainty, have also been used \citep[e.g.][]{LentatiHobson+2014}. 
In cases like the amplitudes of a Fourier series Gaussian process (e.g., \texttt{PowerlawRedNoiseGP}), the priors are defined by the model itself.

Care must be taken to ensure that the data provides enough S/N for the parameter for the `cheat' prior distribution to be valid, lest we effectively do circular analysis \citep{KriegeskorteSimmons+2009}.
A `cheat' prior can also become invalid if the parameter has a physical range, e.g., the sin-inclination of the binary orbit $\sin\iota \in [0,1]$, and the frequentist measurement is close to the physical upper/lower bound. (See Table \ref{tab:sini-priors} for physically motivated priors on different parameterizations of the inclination $\iota$.)

In cases where a physically motivated prior can be analytically derived, \vela{} uses those priors.
For other parameters, `cheat' priors with a user-defined width are used by default.
Crucially, the user may override any of these defaults with the help of univariate distributions defined in \texttt{Distributions.jl}.
The \pyvela{} interface accepts such user-defined univariate prior distributions in the form of a \texttt{JSON} file (see Figure \ref{fig:json-prior} for an example).
Finally, if the prior distributions described above are inadequate, the user also has the flexibility of defining their own prior distributions outside the \vela{} framework since \vela{} is not closely coupled to any sampler.

\begin{table*}[tp]
    \centering
    \begin{tabular}{ccccccc}
\hline
\textbf{Parameter } & \textbf{Binary} & \multirow{2}{*}{\textbf{Definition}} & \multicolumn{4}{c|}{\textbf{Prior distribution}}\tabularnewline
\cline{4-7}
\textbf{name} & \textbf{Models} &  & \textbf{Support} & \textbf{PDF} & \textbf{CDF} & \textbf{Quantile}\tabularnewline
\hline
\hline
KIN & BinaryDDK & Orbital inclination ($\iota$) & $[0,\pi/2]$ & $\sin\iota$ & $1-\cos\iota$ & $\arccos\left(1-q\right)$\tabularnewline
\hline
\multirow{3}{*}{SINI} & BinaryDD & \multirow{2}{*}{$s=\sin\iota$} & \multirow{3}{*}{$[0,1]$} & \multirow{3}{*}{$\frac{s}{\sqrt{1-s^{2}}}$} & \multirow{3}{*}{$1-\sqrt{1-s^{2}}$} & \multirow{3}{*}{$\sqrt{\left(2-q\right)q}$}\tabularnewline
 & BinaryELL1 &  &  &  &  & \tabularnewline
 & BinaryELL1k &  &  &  &  & \tabularnewline
\hline
\multirow{2}{*}{STIGMA} & BinaryDDH & \multirow{2}{*}{$\varsigma=\frac{\sin\iota}{1+\cos\iota}$} & \multirow{2}{*}{$[0,1]$} & \multirow{2}{*}{$\frac{4\varsigma}{\left(1+\varsigma^{2}\right)^{2}}$} & \multirow{2}{*}{$\frac{2\varsigma^{2}}{1+\varsigma^{2}}$} & \multirow{2}{*}{$\sqrt{\frac{q}{2-q}}$}\tabularnewline
 & BinaryELL1H &  &  &  &  & \tabularnewline
\hline
SHAPMAX & BinaryDDS & $S=-\ln(1-\sin\iota)$ & $[0,\infty)$ & $\frac{1-e^{-S}}{\sqrt{2e^{S}-1}}$ & $1-e^{-S}\sqrt{2e^{S}-1}$ & $\ln\left[\frac{\sqrt{2q-q^{2}}+1}{(q-1)^{2}}\right]$\tabularnewline
\hline
\end{tabular}
    \caption{Physically motivated prior distributions for different parameterizations of the orbital inclination $\iota$ assuming that $\cos\iota$ is uniformly distributed in $[0,1]$. The support, probability density function (PDF), cumulative distribution function (CDF), and quantile function (relevant for computing the prior transform) are listed. See Table \ref{tab:components} for the definitions of different binary models.}
    \label{tab:sini-priors}
\end{table*}

\begin{figure*}
\begin{lstlisting}[language=Python]
{
    "EFAC": {
        "distribution": "LogNormal",
        "args": [0.0, 0.5]
    },
    "EQUAD": {
        "distribution": "LogUniform",
        "args": [0.01, 2.0]
    }
    "M2": {
        "distribution": "Normal",
        "args": [1.0, 0.02],
        "lower": 0.0
    }
}
\end{lstlisting}
\caption{An example of the user-defined prior distributions represented in the \texttt{JSON} format. `\texttt{distribution}' should be a distribution available in the \texttt{Distributions.jl} package \citep{BesançonPapamarkou+2021}.
`\texttt{args}' contains the arguments to construct a distribution object.
The `\texttt{upper}' and `\texttt{lower}' options allow the distribution to be truncated.
Only univariate distributions with constant hyperparameters are supported.
The arguments and the upper and lower bounds have the same units used in the \parf{} files where applicable.
For example, the prior distribution for M2 above corresponds to a normal distribution with mean 1 $M_{\odot}$ and standard deviation 0.02 $M_{\odot}$ truncated to exclude negative values.
}
\label{fig:json-prior}
\end{figure*}

\section{Representation of ECORR}
\label{sec:ecorr}
In this section we describe the representation of the ECORR noise used in \vela{} following \citet{JohnsonMeyers+2024} and \citet{SusobhananKaplan+2024}\footnote{Please note that there are typos in equations (21-23) of \citet{SusobhananKaplan+2024} where factors of $c_a^2$ are missing. The expressions below show the correct versions.}.
When ECORR is the only correlated contribution to $\textbf{C}$ (i.e., the time-correlated noise components, if any, are included as a delay or phase correction rather than in $\textbf{C}$), assuming that there are no overlapping ECORR groups, $\textbf{C}$  can be written as
\begin{align}
    \textbf{C} = \textbf{N} + \sum_{ab} c_a \textbf{v}_{ab} \textbf{v}_{ab}^T\,,
\end{align}
where $c_a^2$ is the ECORR weight, the index $a$ represents the different systems for which the ECORRs are assigned, $b$ represents different observing epochs, and $\textbf{N}$ is a diagonal matrix. 
The $\textbf{v}_{ab}$ are vectors that have $1$s for TOAs belonging to the system $a$ and epoch $b$, and $0$s otherwise.
It is clear that $\textbf{C}$ is block-diagonal in this case, and each block $\textbf{C}_{ab}$ can be written as
\begin{align}
    \textbf{C}_{ab} = \textbf{N}_{ab} + c_a \textbf{v}_{ab} \textbf{v}_{ab}^T\,,
\end{align}
where $\textbf{N}_{ab}$ is the portion of $\textbf{N}$ corresponding to the system $a$ and epoch $b$.
The inverse and determinant of $\textbf{C}_{ab}$ can be written as 
\begin{align}
    \textbf{C}_{ab}^{-1} &= \textbf{N}_{ab}^{-1} - \frac{c_a^2 \textbf{N}_{ab}^{-1} \textbf{v}_{ab} \textbf{v}_{ab}^T  \textbf{N}_{ab}^{-1}}{1 + c_a^2 \textbf{v}_{ab}^T  \textbf{N}_{ab}^{-1}\textbf{v}_{ab}}\,,\\ 
    \det \textbf{C}_{ab} &=  \det \textbf{N}_{ab} \times \left(1 + c_a^2 \textbf{v}_{ab}^T  \textbf{N}_{ab}^{-1}\textbf{v}_{ab}\right)\,.
\end{align}
Defining an inner product $\left(\textbf{x}_{ab}|\textbf{y}_{ab}\right)=\textbf{x}_{ab}^T \textbf{N}_{ab}^{-1}\textbf{y}_{ab}$ and writing the log-likelihood function as a sum $\ln L=\sum_{ab}\ln L_{ab}$, we can write
\begin{align}
    \ln L_{ab} = -\frac{1}{2}\left\{
        \left(\textbf{r}_{ab}|\textbf{r}_{ab}\right)
        + \ln \det \textbf{N}_{ab}
        - \frac{c_a^2 \left(\textbf{r}_{ab}|\textbf{v}_{ab}\right)^2}{1 + c_{a}^2 \left(\textbf{v}_{ab}|\textbf{v}_{ab}\right)}
        + \ln\left[1 + c_{a}^2 \left(\textbf{v}_{ab}|\textbf{v}_{ab}\right)\right]
    \right\}\,.
    \label{eq:ln_Lab}
\end{align}

Since $\textbf{N}_{ab}$ is diagonal, the inner products appearing in equation \eqref{eq:ln_Lab} can be evaluated with linear time complexity with a single pass over the TOAs without the need for any dynamic memory allocations.
Further, it is straightforward to see that the evaluation of $\ln L$ can be trivially parallelized over the ECORR groups $ab$.

\section{Representation of red noise processes}
\label{sec:rednoise}
We represent the red noise processes affecting the TOAs as delays represented by truncated Fourier series
\begin{align}
    \Delta_{\text{RN};\alpha}(t) = \left(\frac{\nu_{\text{ref}}}{\nu}\right)^\alpha \sum_{j=0}^{N_{\text{harm}} } \left[ 
        a_j \cos(2\pi j f_1 (t-t_0)) + b_j \sin(2\pi j f_1 (t-t_0))
    \right]\,,
\end{align}
where $f_1$ is a fundamental frequency, $\alpha$ is the chromatic index, $\alpha=0$ represents spin noise, and $\alpha=2$ represents dispersion noise.
$f_1$ is usually taken to be the reciprocal of the total observation span of the dataset.

We provide two types of representation for red noise. 
\texttt{WaveX} (spin noise), \texttt{DMWaveX} (dispersion noise), and \texttt{CMWaveX} (variable-$\alpha$ chromatic noise) treat the coefficients $a_j$ and $b_j$ as unconstrained free parameters with uninformative priors.
This representation is useful for reconstructing cross-pulsar correlations from single-pulsar noise analysis runs post facto \citep{ValtolinaVan_Haasteren2024}. 
(This will be explored in a future work.)

The Gaussian process models \texttt{PowerlawRedNoiseGP}, \texttt{PowerlawDispersionNoiseGP}, and \texttt{PowerlawChromaticNoiseGP} impose Gaussian prior distributions on the Fourier coefficients such that 
$\left\langle a_j\right\rangle = \left\langle b_j\right\rangle=0$, 
$\left\langle a_j a_k\right\rangle = \left\langle b_j b_k\right\rangle=\sigma_j^2 \delta_{jk}$, 
$\left\langle a_j b_k\right\rangle = 0$.
Further, the spectral power densities $\sigma_j$ follow a power law spectrum 
\begin{align}
    \sigma_j^2 = \frac{A^2}{12\pi^2 f_{\text{yr}}^3} f_1 \left(\frac{f_{\text{yr}}}{jf_1}\right)^\gamma\,.
\end{align}
\updated{Note that certain physical models of the spin noise introduce a low-frequency turnover in its spectrum \citep{GoncharovZhuThrane2020}.
While this power spectral model is currently unavailable, it will be included in a future version of \vela{}.}

It turns out that the joint prior distribution of $a_j$ or $b_j$ and the power law parameters $A$ and $\gamma$ displays Neal's funnel-like geometry \citep{Neal2003}, which is hard for MCMC samplers to explore. 
We handle this by treating $\bar{a}_j=a_j/\sigma_j$ and $\bar{b}_j=b_j/\sigma_j$ as free parameters rather than $a_j$ and $b_j$.
It is easy to see that these new parameters are a priori unit-normal distributed.
This also has the advantage of simplifying the implementation of prior transform functions by ensuring that the prior distribution of each parameter is independent of the other parameters.

Note that these Gaussian process models have $2N_{\text{harm}}+2$ number of parameters.
This leads to high-dimensional parameter spaces which can be challenging to sample. 
On the other hand, treating the time-correlated noise processes as delays has the advantage that the log-likelihood can be evaluated in linear time without memory allocations (see Appendix \ref{sec:ecorr}).
The sampling challenges posed by the large number of Fourier coefficients can be addressed somewhat by employing Gibbs sampling for those parameters, e.g., \citet{LaalLamb+2023}, since it turns out that the conditional distributions for these parameters can be analytically derived.  
This will be explored in a future work.


\bibliography{vela-overview}{}
\bibliographystyle{aasjournal}



\end{document}